\documentclass[twoside,fleqn]{article}
\usepackage{espcrc2}
\usepackage{graphicx}
\newcommand{\bmp}{\mathbf p}
\def\lsim{\mathrel{\raise.3ex\hbox{$<$\kern-.75em\lower1ex\hbox{$\sim$}}}}
\def\gsim{\mathrel{\raise.3ex\hbox{$>$\kern-.75em\lower1ex\hbox{$\sim$}}}}

\newcommand{\AmS}{{\protect\the\textfont2
  A\kern-.1667em\lower.5ex\hbox{M}\kern-.125emS}}

\title{Implementing invariant mass cuts and finite lifetime effects 
in top-antitop production at threshold}

\author{A.~Hoang$^a$, C.~Rei\ss er$^{a,b}$ and P.~Ruiz-Femen\'\i a$^c$\thanks{I would 
        like to thank S.Narison and his crew for the pleasant atmosphere 
        during the conference. This work is partly supported by the DFG, SFB/TR9. Preprint MPP-2008-137, 
        SFB/CPP-08-84, PITHA 08/28.}
\\ [0.4cm]
$^a$
Max-Planck-Institut f\"ur Physik,
F\"ohringer Ring~6, D--80805 M\"unchen, Germany\\
$^b$
Institut f\"ur Theoretische Teilchenphysik, Universit\"at Karlsruhe, %\\
D-76128 Karlsruhe, Germany \\
$^c$
Institut f\"ur Theoretische Physik E, RWTH Aachen, %\\ 
D--52056 Aachen, Germany%
       }

\begin{document}

\begin{abstract}
The effects of the finite top quark width in the top pair production cross
section close to the threshold are discussed in this talk.
We introduce a $t\bar{t}$ cross section with a cut on the invariant masses
of the top and antitop that can be calculated theoretically with 
effective field theory (EFT) methods.
The matching procedure to implement the physical phase-space boundaries in the NRQCD
framework (``phase-space matching'') is briefly outlined.
\end{abstract}

\maketitle

%\section{INTRODUCTION}

The measurement of the $\sigma(e^+e^-\to t\bar{t})$ line shape at a
linear collider (LC) operating at energies around the top-antitop threshold
($\sqrt{q^2}\sim 350\,$GeV) would allow for very accurate determinations
of the mass, the width and the couplings of the top quark~\cite{Martinez:2002st}.
In view of the accuracy
obtainable at the LC the theoretical uncertainties for the
cross section predictions should be lowered to a level of a few percent~\cite{Hoang:2006pc}.

Close to threshold the top quark pairs are produced with small velocities $v\ll 1$.
In the standard QCD perturbative expansion singular terms $\sim \alpha_s/v,\,\alpha_s\log v$
arise from the ratios of the physical scales involved: the top mass $m_t$, the relative
three-momentum $\bmp\sim m_t v$ and the nonrelativistic energy $E\sim m_t v^2$. 
The summation of these terms can be achieved systematically by means of 
nonrelativistic QCD (NRQCD), the low-energy EFT that describes the quantum
fluctuations of full QCD for the kinematic situation of top quarks close to threshold.
Renormalization group improved calculations of the $t\bar{t}$ total cross section
with NNLL order accuracy for the QCD effects
have become available in this framework~\cite{Hoang:2000ib}.

%\section{FINITE WIDTH EFFECTS}

A study of electroweak corrections in $t\bar{t}$ production at threshold beyond
leading order has not been systematically undertaken until 
recently~\cite{Hoang:2004tg,Hoang:2006pd}. 
The large top
width, being of the same order as the nonrelativistic energy,
is essential in the description of the $t\bar{t}$ threshold dynamics. The dominant
top decay mode in the Standard Model is $t\to b W^+$, which gives $\Gamma_t\sim 1.5\,$GeV.
It was shown~\cite{Fadin} that in the nonrelativistic limit the top-quark
width can be consistently implemented by the replacement
$E\to E+i\Gamma_t$ in the results for the total cross section 
for stable top quarks. This replacement rule accounts for the LL
electroweak corrections to the total $t\bar{t}$ cross section,
but a complete treatment at NLL and NNLL requires the use of an extended NRQCD
formalism accounting for the top quark instability.

\begin{figure}[t]
  \begin{center}
%  \vspace*{-0.1cm}
  \includegraphics[width=0.5\textwidth]{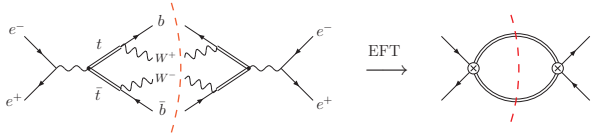}
  \vspace*{-1.3cm}
  \caption{Tree-level double-resonant diagrams in the full and effective
  theories. The dashed line means we extract the imaginary part or, equivalently, that
  we perform the phase space integration along the cut lines.}
  \label{fig1}
  \end{center}
  \vspace*{-0.7cm}
\end{figure}

In the NRQCD framework the quark width effects are incorporated at leading order
through the bilinear quark operators
\begin{equation}
\delta{\cal L} = \sum_{\bmp} \psi_\bmp^\dagger \, \frac{i}{2}\Gamma_t \, \psi_\bmp
+ \sum_{\bmp} \chi_{\bmp}^\dagger \, \frac{i}{2} \Gamma_t \, \chi_\bmp 
\,.
\label{eq:bilinear}
\end{equation}
They represent a LL effect as compared to the typical energy of a Coulombic top quark, of the order of a few GeV. The terms in Eq.~(\ref{eq:bilinear}) describe an absorptive on-shell process that has been integrated out from the theory. In this way, the EFT does not contain any details and
differential information of the decay, but can predict the total cross section, an inclusive observable.
We can use the optical theorem to compute
the total cross section from the imaginary part of the $e^+e^-$ forward scattering amplitude (see Fig~\ref{fig1}). 
The coefficients in Eq.~(\ref{eq:bilinear}) are determined 
from the matching procedure of the EFT to the full theory
(QCD and the electroweak theory).
The contributions from real $b W$ final states are thus included 
through matching conditions of the EFT.
The LO bilinear quark terms in the NRQCD Lagrangian lead to the propagator 
\begin{equation}
\frac{i}{p_0-\bmp^2/(2m_t)+i\Gamma_t/2} 
\label{eq:prop}
\end{equation}
for a top/antitop with momentum $(p_0, \bmp )\sim (m_t v^2,m_t v)$.
The inclusion of a finite width in the top propagator modifies the high-energy behaviour of the phase-space integrations in the EFT. Let us illustrate
this with an example. Consider the double-resonant diagram of Fig.~\ref{fig1} in the full theory with center of mass momentum $q=(2m_t+E,\mathbf{0})$.
When the quark lines are close to their mass shell the 4-particle phase space integration reduces to 
\begin{equation}
\sigma_{t\bar{t}} \sim \int_{-\infty}^{+\infty} \!\!\!\!\!\!\!\! dp_0 
\int_{0}^{+\infty} \!\!\!\!\!\!\!\! d|\bmp| \bmp^2 
\frac{\Gamma_t^2}
{\left| t_1 + i \epsilon \right|^2
 \left| t_2 + i \epsilon \right|^2 }% \sim \frac{\#}{m_t^2}\,v
\label{eq:tree}
\end{equation}
where the variables $t_{1,2}\equiv t_{1,2}(p_0,\bmp)=2m_t ( \frac{E}{2}\pm p_0-\frac{\bmp^2}{2m_t} )$ retain the leading order
terms in the nonrelativistic expansion of the top and antitop off-shellness. The $\Gamma_t^2$
factor in the numerator arises from the phase-space integration of the $b W^+$ and $\bar{b}W^-$ subsystems because in the nonrelativistic limit the invariant masses
of the top/antitop quark can be set to $m_t^2$. 

We notice that though the phase-space in the full theory is cut-off by the top mass,
the nonrelativistic limit formally makes $m_t\to\infty$ and the phase-space boundaries are thus taken to infinity\footnote{Note that the arguments
of the delta and step functions in the phase-space also have to be expanded out in the spirit 
of the threshold expansion.}. The corresponding NRQCD amplitude (diagram on the right in Fig.~\ref{fig1})  reproduces Eq.~(\ref{eq:tree})
when $\Gamma_t$ is treated as an insertion. The NRQCD powercounting, however, tells that the $\Gamma_t \sim E$
term needs to be resummed as part of the top propagator
(see Eq.(2)), thus effectively replacing replacing $i\epsilon \to i\Gamma_t$,
to finally give $\sigma_{t\bar{t}} \propto v/m_t^2$, with 
$v=\sqrt{(E+i\Gamma_t)/m_t}$.

\begin{figure}[t]
  \begin{center}
  \vspace*{-0.1cm}
  \includegraphics[width=0.35\textwidth]{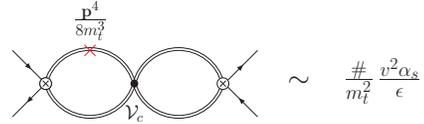}
  \vspace*{-0.5cm}
  \caption{$(e^+e^-)(e^+e^-)$ scattering amplitude with a kinetic energy correction
and a Coulomb potential. Its imaginary part gives a NNLO divergent contribution to $\sigma_{t\bar{t}}$.}
  \label{fig2}
  \end{center}
  \vspace*{-0.7cm}
\end{figure}

Despite the integration limits, the tree-level integration in Eq.~(\ref{eq:tree}) is finite.
However the integrand becomes more sensitive to large momentum regions once
we include relativistic corrections $\sim \bmp^2/m_t^2$. Using dimensional regularization
these subleading contributions can lead to $1/\epsilon$ singularities
if the high energy behaviour
of the EFT phase-space integration is logarithmic. This occurs for some of the NNLO 
matrix elements containing a single insertion of the Coulomb potential and a $\bmp^2/m_t^2$ correction.
Figure~\ref{fig2} shows an example.
Similar UV divergences had also been noted in Ref.~\cite{Hoang:2004tg},
where
the imaginary parts of the NRQCD matching conditions at NNLL
coming from $Wb$ cuts in the full theory were determined and inserted
into NRQCD matrix elements. These UV divergences would not exist for
a stable top quark. These {\it phase space divergences} can be
handled with the usual EFT renormalization 
techniques. In this specific example
the UV divergences are absorbed by
$(e^+e^-)(e^+e^-)$ operators that will
thus have complex Wilson coefficients
$\tilde{C}_{V/A}(\mu)$~\cite{Hoang:2004tg}.
The imaginary parts of these Wilson coefficients
contribute to the $t\bar{t}$ cross section through the optical theorem,
$\Delta \sigma_{\bar{t}t}=\mathrm{Im} [ \tilde{C}_V + \tilde{C}_A ]$. The renormalization
procedure of course does not determine the matching conditions at the hard scale, $\tilde{C}_{V/A}(\mu=m_t)$,
and a matching prescription is required in order to put back into the EFT the information about the
physical phase-space boundaries. A possible approach is to define the matching condition
as the difference between the full theory computation for the $b W^+\bar{b}W^-$ 
total cross section (considering $b$ and $W$ as stable) and the NRQCD result, similarly to the procedure
used in Ref.~\cite{Beneke:2007zg} for the case of $W^+W^-$ threshold production. 
Obviously, the full theory computation with the 4-particle final state is a rather complicated task 
beyond tree-level, even if an asymptotic expansion technique is employed.

\begin{figure}[t]
  \begin{center}
  %\vspace*{-0.1cm}
  \includegraphics[width=0.3\textwidth]{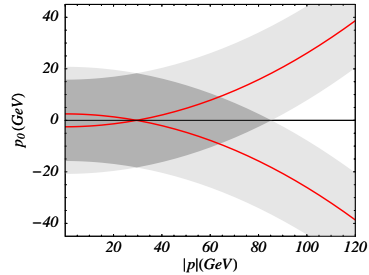}
  \vspace*{-0.8cm}
  \caption{Phase-space integration region spanned in the $(p_0,\bmp)$ variables. We
  have chosen $E=5\,$GeV and $\Lambda=80\,$GeV for this example.}
  \label{fig3}
  \end{center}
  \vspace*{-.7cm}
\end{figure}

At this point,
a different path can be followed, that will reduce the amount of full theory input required and 
allow us to make contact with the experimentally measured cross section. Among the
experimental cuts that need to be applied to select $b W^+\bar{b}W^-$ events in
$t\bar{t}$ production, cuts on the top and antitop invariant masses will be essential.
Close to the threshold, it is more natural to define the cut on the
the top invariant mass squared $p_{t}^2$
with respect to its on-shell value, {\it i.e.} 
\begin{equation}
|p_{t}^2-m_t^2| \le \Lambda^2\,,
\label{eq:lambda}
\end{equation}
and similarly for the antitop. In the nonrelativistic limit, the top and
antitop off-shellness are characterized
by the variables $t_{1,2}(p_0,\bmp)$. The cutoff $\Lambda$
divides the phase-space into three different regions: the double-resonant region, 
with both $|t_{1,2}|<\Lambda^2$, the single-resonant regions,
corresponding to ($|t_1|<\Lambda^2$, $|t_2|>\Lambda^2$) 
or ($|t_2|<\Lambda^2$, $|t_1|>\Lambda^2$), and the non-resonant region, 
where both $|t_{1,2}|>\Lambda^2$. The regions are shown in the phase-space
diagram in Fig.~\ref{fig3} (gray, light-gray and white, respectively). 
The red lines in Fig.~\ref{fig3} correspond to the top and antitop on-shell conditions $t_{1,2}=0$. 
For the case $\Gamma_t=0$, the cut through the top and antitop
EFT propagators yields a delta function instead of a Breit-Wigner distribution, and the phase-space shrinks to the point given by the intersection of the two on-shell lines ($|\bmp|=\sqrt{mE}$). 
Restricting the phase-space boundaries to the double-resonant region 
defines a new threshold $t\bar{t}$ cross section, $\sigma_{t\bar{t}}(\Lambda)$,
that corresponds to taking into account only those measured $bW^+$ and $bW^-$ events
with invariant masses inside the window given by the condition (\ref{eq:lambda}).
The cutoff captures the double-resonant region where the top and antitop
are close to their mass-shell and the NRQCD expansion applies.
It excludes those parts of the phase space that are not described
well by the EFT.
The latter is true provided that the value of $\Lambda$ 
formally satisfies $(E,\Gamma)\ll \Lambda^2/m_t \ll m_t$. 
The physical meaning of the new matching
condition, 
\begin{equation}
\mathrm{Im}\,\tilde{C}_{V/A} \doteq \sigma_{t\bar{t}}(\Lambda)-\sigma_{t\bar{t}}^{\mathrm{EFT}}\,,
\label{eq:matching}
\end{equation}
becomes thus clear in this approach: it corrects for the wrong behaviour
of the effective propagators outside the nonrelativistic domain 
($\sigma_{t\bar{t}}^{\mathrm{EFT}}$ is the standard NRQCD result, {\it i.e.} for $\Lambda\to\infty$).
We can go a step further by realizing that inside the double-resonant region
the full theory cutoff cross section $\sigma_{t\bar{t}}(\Lambda)$ can be 
nonrelativistically expanded since $\Lambda\ll m_t$. The computation of
 $\sigma_{t\bar{t}}(\Lambda)$ can thus be carried out using NRQCD
Feynman rules and, in this way,
the whole approach is interpreted as a matching procedure within the EFT itself.
There are of course contributions with the same $bW^+ \bar{b}W^-$ final state
that are not described by the EFT computation (the interferences of
amplitudes containing a single top/antitop line, for example).
The latter are part of the irreducible background, whose numerical contribution
to $\sigma_{t\bar{t}}(\Lambda)$ can be estimated using the tree-level predictions since
there is no Coulomb-enhancement for these terms at higher orders in $\alpha_s$.

The coefficients $\tilde{C}_{V/A}$ provide the corrections to the 
standard NRQCD result due to phase-space effects at a practical level.
From a formal point of view, we have overlooked the fact that the corrections $\tilde{C}_{V/A}$ defined through Eq.~(\ref{eq:matching}) are
energy dependent, and cannot be cast as the Wilson coefficients of  
local $(e^+e^-)(e^+e^-)$ operators. A local action is needed in order 
to use the optical theorem to determine the cross section 
from the imaginary part of the $e^+e^-$ forward scattering amplitude.
The natural solution is to write the corrections $\tilde{C}_{V/A}$ as the
coefficients of an operator product expansion in $(e^+e^-)(e^+e^-)$ operators with
increasing powers of time derivatives $i\partial_0\sim E$. This is achieved practically
by expanding the phase-space effects in powers of $E/\Lambda$.
For example, the matching of the ${\cal O}(\alpha_s^0)$ diagram shown
schematically in Fig.~\ref{fig4} yields the phase-space correction:
\begin{figure}[t]
%  \begin{center}
  \vspace*{-0.1cm}
  \includegraphics[width=0.4\textwidth]{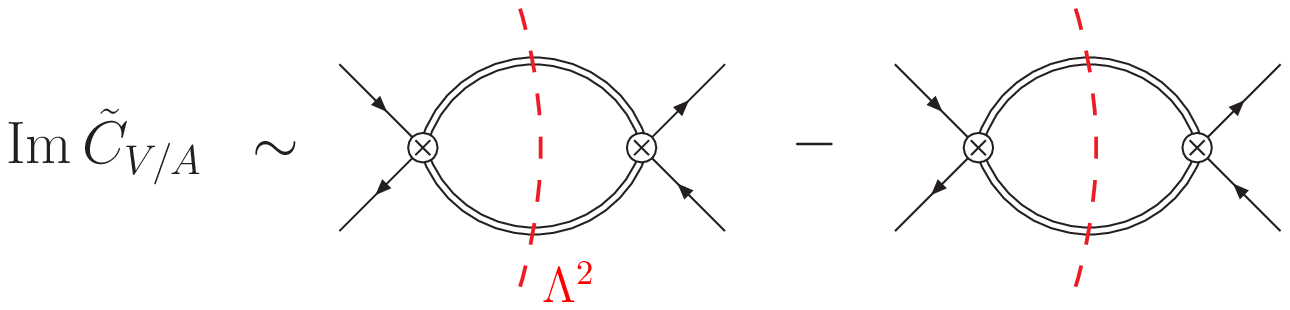}
  \vspace*{-0.8cm}
  \caption{${\cal O}(\alpha_s^0)$ phase-space matching.}
  \label{fig4}
%  \end{center}
  \vspace*{-0.4cm}
\end{figure}
\begin{equation}
\sum_n \Big(\frac{E}{m_t}\Big)^n \tilde{C}^{(n)}_{V/A}
\sim  \frac{i}{m_t^2} \frac{\Gamma_t}{\Lambda}
 \Big( 1 + \frac{m_t E}{3\Lambda^2}+\dots \Big)
\label{eq:born}
\end{equation}
As expected, the result vanishes for $\Gamma_t\to 0$. We also note that
the non-local contribution $\propto v$ in the diagrams
of Fig.~\ref{fig4} cancels in the difference. The term proportional to
$\Gamma_t/\Lambda$ gives the dominant phase-space correction to the $t\bar{t}$ cross
section, whose relevance will depend on the chosen value for the cutoff. We mentioned 
above that our approach is formally valid for $\Lambda^2/m_t\ll m_t$. This condition may seem
necessary due to the appearance of powers of $\Lambda^2/m_t^2$
when higher dimensional operators are considered.
However, a numerical analysis of the coefficients of such contributions
shows that the OPE series has a good convergence even for values 
of $\Lambda$ close to $m_t$, although smaller values ($\Lambda\lsim 0.6 m_t$) 
are required to lower their size below the desired accuracy of 1\%. We shall
use the counting $\Lambda\sim m_t$ to identify the order at which
the phase-space matching corrections contribute. In this scheme, the
correction  from Eq.~(\ref{eq:born}) scales as $\Gamma/\Lambda \sim v^2$, and thus represents a NLO correction.

The NRQCD calculation of the $t\bar{t}$ cross section sums up the insertions of Coulomb
potentials to all orders in $\alpha_s$. Since the suggested OPE expansion of
the phase-space corrections necessarily involves a matching of diagrams
with a given order in $v$ and $\alpha_s$, we have to test also the convergence of the matching corrections arising from the addition of Coulomb potentials. The insertion of a Coulomb potential yields a factor $\alpha_s/v$ for $\Lambda\to \infty$; therefore we shall expect a suppression factor $\alpha_s m_t/\Lambda$ for every Coulomb exchange added to the matching relation of Fig.~\ref{fig4}. An optimal range of values for the cutoff is
to be determined in order to find the required suppression of both the $(\Lambda^2/m_t^2)^k$ power-counting breaking terms and the $(\alpha_s m_t/\Lambda)^k$ Coulomb phase-space corrections. Let us note that the phase-space matching of the $(e^+e^-)(e^+e^-)$ operators at 2-loops requires the removal of $\Lambda$-dependent non-local terms 
by 1-loop diagrams with phase-space subtractions in the production 
currents, in analogy to the process of removal of subdivergences in the usual 
renormalization procedure.

Results for all the phase-space corrections needed for a determination
of the threshold $t\bar{t}$ cross section with NNLL accuracy 
and a complete numerical analysis shall be presented in a forthcoming 
publication.%~\cite{our}.

\end{document}